\documentclass[aps,pra,showpacs,twocolumn,amsmath,amssymb]{revtex4}

\usepackage{graphicx}
\usepackage{amsmath}
\usepackage{color}
\usepackage{dcolumn}
\usepackage{bm}
\usepackage{booktabs}

\begin{document}

\title{Scattering lengths of calcium and barium isotopes}

\author{U.\ Dammalapati}
\author{L.\ Willmann}
\email[]{l.willmann@rug.nl}\affiliation{Kernfysisch Versneller
Instituut (KVI), University of Groningen, Zernikelaan 25, 9747 AA
Groningen, The Netherlands.}
\author{S.\ Knoop}
\email[]{s.knoop@vu.nl} \affiliation{LaserLaB Vrije Universiteit,
De Boelelaan 1081, 1081 HV Amsterdam, The Netherlands.}
\date{\today}

\begin{abstract}
We have calculated the $s$-wave scattering length of all the even
isotopes of calcium (Ca) and barium (Ba), in order to investigate
the prospect of Bose-Einstein condensation (BEC). For Ca we have
used an accurate molecular potential based on detailed
spectroscopic data. Our calculations show that Ca does not provide
other isotopes alternative to the recently Bose condensed
$^{40}$Ca that suffers strong losses because of a very large
scattering length. For Ba we show by using a model potential that
the even isotopes cover a broad range of scattering lengths,
opening the possibility of BEC for at least one of the isotopes.
\end{abstract}


\maketitle

\section{Introduction}

Knowledge of the $s$-wave scattering length plays a crucial role
in the achievement of Bose-Einstein condensation (BEC) in
ultracold atomic gases~\cite{BURNETT2002}. A positive sign of the
scattering length indicates that a Bose-Einstein condensate is
stable whereas a negative sign indicates an unstable or collapsing
condensate. Also its magnitude is crucial in the formation of BEC,
as it determines the elastic and inelastic collision rates. For
efficient evaporative cooling a large scattering length is
required, however for a too large scattering length three-body
recombination loss limits the formation and collisional stability
of a BEC. This gives rise to a range of about $50-200a_0$
($a_0$=0.5292~nm) that can be considered favorable.

Alkaline-earth elements and the alike ytterbium (Yb) system have
been the center of attraction in ultra-cold atom research because
of their unique atomic structure and the availability of different
stable isotopes with zero nuclear spin (bosons) and with large
nuclear spin (fermions). They are being used in frequency
metrology~\cite{LUDLOW2008,LEMKE2009,DEREVIANKO2011} and are
proposed for quantum information processing~\cite{DALEY2008}.
Recently, BEC has been achieved in $^{40}$Ca~\cite{KRAFT2009},
$^{84}$Sr~\cite{STELLMER2009,MARTINEZ2009},
$^{86}$Sr~\cite{STELLMER2010} and $^{88}$Sr~\cite{MICKELSON2010}.
Also BEC of different Yb isotopes has been reported:
$^{174}$Yb~\cite{TAKASU2003}, $^{170}$Yb~\cite{FUKUHARA2007},
$^{176}$Yb~\cite{FUKUHARA2009} and $^{168}$Yb~\cite{SUGAWA2011},
of which the last one has an abundance of only 0.13\%,
highlighting the possibility of using rare isotopes to achieve
BEC.

In general, isotopes of an element have different scattering
lengths. Therefore the ability of BEC formation depends crucially
on the chosen isotope. For instance, the scattering length of
$^{84}$Sr is 123$a_{0}$~\cite{STEIN2010}, which made it an ideal
candidate to achieve BEC despite its low abundance of only
0.6\%~\cite{ZELEVINSKY2009}, as compared to the more abundant
$^{86}$Sr (10\%, 800$a_{0}$ \cite{STEIN2010}) and $^{88}$Sr (83\%,
-2$a_{0}$ \cite{STEIN2010}), either suffering strong
losses~\cite{STELLMER2010} or requiring sympathetic cooling with
another isotope~\cite{MICKELSON2010}, respectively. The $^{40}$Ca
isotope has a large scattering length of about
440$a_{0}$~\cite{KRAFT2009}, limiting the size and stability of
BEC.

Motivated by the success of BEC in Sr and Yb even isotopes, and
$^{40}$Ca, we carried out calculations to obtain the scattering
lengths for all the even (bosonic) isotopes of Ca and Ba, for
which mass and abundance are summarized in Table~\ref{isotopes}.
Laser cooling and trapping of all stable Ca isotopes have been
reported~\cite{HOEKSTRA2005,DAMMALAPATI2010}. The demonstration of
magneto-optical trapping of Ba~\cite{DE2009} has opened its use
for ultra-cold collision studies, photoassociation spectroscopy
and Bose-Einstein condensation. Recently, an optical clock based
on ultracold Ba has been proposed~\cite{HUA2011}.

\begin{table}[b]
\caption{Mass and abundance of all stable even isotopes of Ca and
Ba \cite{SANSONETTI2005}.} \label{isotopes}
\begin{ruledtabular}
\begin{tabular}{lccccc}
\multicolumn{6}{c}{Calcium} \\
\hline
  Isotope   & 40      & 42            & 44     & 46 & 48     \\
  \hline
    Mass (amu) & 39.963    & 41.959     &43.955      & 45.954    & 47.953   \\
   Abundance (\%) & 96.941    & 0.647      &2.086       &  0.004  & 0.187      \\
 \hline
 \multicolumn{6}{c}{Barium} \\
\hline
  Isotope   & 130      & 132      & 134     & 136    & 138    \\
  \hline
    Mass (amu) & 129.906   & 131.905  &  133.904   &135.905      & 137.905      \\
   Abundance (\%)& 0.11     & 0.10   &  2.42  &7.85      & 71.70        \\
\end{tabular}
\end{ruledtabular}
\end{table}

\section{Theory}

The scattering properties of ground state atoms are obtained from
the underlying two-body ground state potentials. For
alkaline-earth atoms there is only one molecular ground state
potential, namely the singlet $X^1\Sigma^+_g$ potential.
Furthermore, the lack nuclear spin and therefore hyperfine
structure for the even (bosonic) isotopes highly reduces the
number of collision channels compared with the alkali metal
systems.

\begin{figure}[t!]
\center
\includegraphics[width = 85 mm, angle = 0]{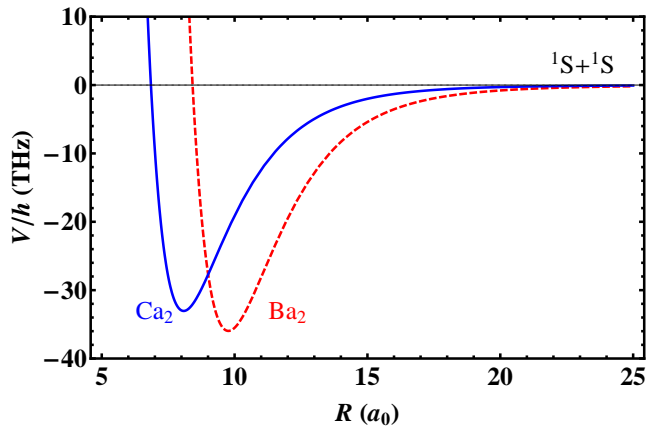}\\
\caption{(Color online) The $X^1\Sigma^+_g$ potentials of Ca$_2$
and Ba$_2$ as a function of internuclear distance. The solid blue
curve is the potential of Ca$_2$ taken from
Ref.~\cite{ALLARD2003}. The dashed red curve is the Tang-Toennies
potential of Ba$_2$ taken from
Ref.~\cite{LI2011}.}\label{potentials}
\end{figure}

To calculate the $s$-wave scattering length we solve the 1D radial
Schr\"{o}dinger equation with zero angular momentum and vanishing
kinetic energy,
\begin{equation}\label{Schrodinger}
\psi''(R)+\frac{2\mu}{\hbar^2}\left[E-V(R)\right]\psi(R)=0,
\end{equation}
where $\mu$ is the reduced mass, $R$ the internuclear distance,
$V(R)$ the Born-Oppenheimer potential (here the $X^1\Sigma^+_g$
potential), and $E$ the kinetic energy (below $\mu$K). The
asymptotic form of the wavefunction is
\begin{math}\psi(R)\propto\sin[k(R-a)]\end{math}, where $k=\sqrt{2\mu
E/\hbar^2}$, and $a$ is the scattering length. We fit the
asymptotic form to the solution of Eq.~(\ref{Schrodinger}) for
large $R$ in order to obtain $a$. Within the Born-Oppenheimer
approximation, the potentials are identical for all the isotopes
of a particular atomic system. Therefore, for a given potential
$V(R)$, one only needs to adjust $\mu$ to obtain $a$ for all the
isotopes.

To obtain reliable scattering lengths accurate potentials are
needed. With the exception of systems with only a few electrons,
like metastable helium~\cite{PRZYBYTEK2005}, \textit{ab initio}
potentials are in general not accurate enough and constraints from
experimental data are required. For Ca$_2$ an analytic
representation of the $X^1\Sigma^+_g$ potential based on an
extensive set of experimental data from the Tiemann group is
available~\cite{ALLARD2003}. For Ba$_2$ such an accurate potential
is not available. Here we rely on the analytical representation
according to the Tang-Toennies potential model~\cite{LI2011},
which has shown to be able to reproduce accurately the
$X^1\Sigma^+_g$ potentials of Ca$_2$ \cite{YANG2009} and Sr$_2$
\cite{LI2011}. The used Ca$_2$ and Ba$_2$ potentials are shown in
Fig.~\ref{potentials}.

\section{Results}

\subsection{Calcium}\label{calcium}

Information on the Ca$_2$ $X^1\Sigma^+_g$ potential has been
gathered by several spectroscopic methods, including
photoassociation
spectroscopy~\cite{ZINNER2000,DEGENHARDT2003,VOGT2007},
Fourier-transform spectroscopy~\cite{ALLARD2002} and filtered
laser excitation technique~\cite{ALLARD2003}. Based on these
detailed spectroscopic data, an interval of the scattering length
for $^{40}$Ca is determined to be $340-700$$a_0$~\cite{VOGT2007}.
In the $^{40}$Ca BEC experiment a scattering length of $a\approx
440$$a_0$ was estimated from a measurement of the chemical
potential \cite{KRAFT2009}, without quoting an uncertainty.
Therefore we take the scattering length range of $340-700$$a_0$ as
starting point of our calculations.

\begin{figure}[t!]
\center
\includegraphics[width = 85 mm, angle = 0]{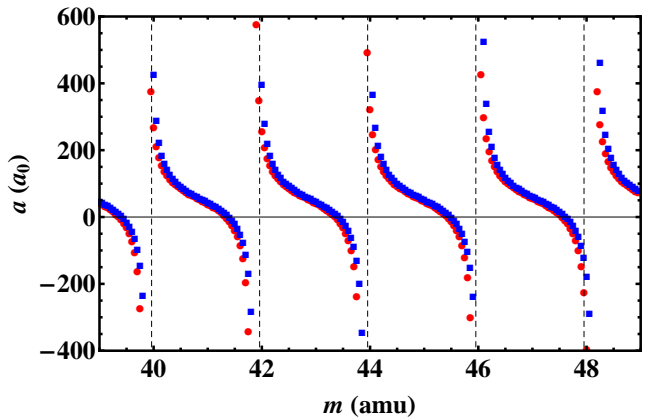}\\
\caption{(Color online) The mass dependence of the scattering
length for Ca, using the $X^1\Sigma^+_g$ potential of
Ref.~\cite{ALLARD2003}, showing calculations with $C_6$
coefficients that give rise to 340$a_0$ (red closed circles) and
700$a_0$ (blue closed squares) for $^{40}$Ca. The plotted mass is
twice the reduced mass and the dashed vertical lines indicate the
mass of the even isotopes of Ca.} \label{Cascattering}
\end{figure}

We have taken the analytical representation and parameter values
of the Ca$_2$ $X^1\Sigma^+_g$ potential from
Ref.~\cite{ALLARD2003} (see Fig.~\ref{potentials}). We allow the
$C_6$ long-range coefficient to vary within a factor 0.991 to
1.003 from its reported value \cite{ALLARD2003}, in order to
reproduce the scattering length interval for $^{40}$Ca. We then
take the mass as a variable parameter and calculate the scattering
length for the full mass interval spanned by the stable even
isotopes. In this way we transfer the knowledge on $^{40}$Ca to
the other isotopes.

The results for Ca are shown in Fig.~\ref{Cascattering}. The
plotted mass is twice the reduced mass, which for homonuclear
collisions simply is the atomic mass. The dashed vertical lines
indicate the masses of the even isotopes, from which the
scattering lengths for homonuclear collisions are directly read
off. Two calculations are shown, obtained from potentials with
$C_6$ coefficients that give rise to 340$a_0$ (red closed circles)
and 700$a_0$ (blue closed squares) for $^{40}$Ca. The scattering
length shows the expected behavior as a function of mass, with a
regular pattern of scattering resonances. They can be understood
from the mass dependence of the vibrational splitting, and
therefore the number of bound states. A scattering resonance
appears at those mass values at which a new vibrational state
becomes bound, i.\ e.\ where the least bound vibrational state has
zero binding energy. We find that accidentally all even isotopes
are located close to such a scattering resonance, giving rise to
large positive or negative scattering lengths.

The scattering length intervals for the different isotopes are
given in Table~\ref{Cascat}. It is clear that Ca does not provide
an isotope with a favorable scattering length. At most one can
state that $^{42}$Ca has a slightly smaller scattering length than
$^{40}$Ca. In addition, the interisotope scattering lengths are
given for collisions between $^{40}$Ca and the other isotopes, and
$^{42}$Ca+$^{44}$Ca. Here we find favorable scattering lengths for
$^{40}$Ca+$^{42}$Ca, $^{40}$Ca+$^{46}$Ca and $^{42}$Ca+$^{44}$Ca.
Taking into consideration the large scattering length of $^{44}$Ca
and $^{46}$Ca, the only interesting mixture in view of sympathetic
cooling is $^{40}$Ca+$^{42}$Ca.

\begin{table}
\caption{Scattering lengths of even Ca isotopes (left), as well as
interisotopic scattering lengths for all combinations with
$^{40}$Ca, and $^{42}$Ca+$^{44}$Ca  (right).} \label{Cascat}
\begin{ruledtabular}
\begin{tabular}{cc|cc}
     &  $a$ ($a_{0}$)    &   &  $a$ ($a_{0}$)\\
  \hline
   40+40            &  $+340\ldots+700$ \cite{VOGT2007}       &   40+42      &  $+45\ldots+50$                       \\
   42+42           &  $+320\ldots+640$                 &    40+44     &  $\pm\infty$        \\
   44+44              &  $+460\ldots+1800$     &  40+46              &  $+62\ldots+68$         \\
   46+46             & $\pm\infty$   &    40+48      &  $-66\ldots-39$   \\
   48+48         & $-230\ldots$$-120$  &   42+44    &          $+47\ldots+52$        \\
\end{tabular}
\end{ruledtabular}
\end{table}

\subsection{Barium}

Knowledge of the Ba$_2$ $X^1\Sigma^+_g$ potential is sparse,
because of a limited amount of experimental data
\cite{CLEMENTS1985,LEBEAULT1998}, which cover only a small part of
the vibrational spectrum. Similarly, theoretical data is scarce.
The binding energies of Ba$_{2}$ were calculated by
Jones~\cite{JONES1979}. First wave function based quantum chemical
calculation was reported by Ref.~\cite{ALLOUCHE1995}, most
recently by Ref.~\cite{MITIN2009}. The most accurate calculations
of the long-range coefficients $C_6$, $C_8$, $C_{10}$ are reported
by Porsev and Derevianko \cite{PORSEV2002,PORSEV2006}.

We have used the Tang-Toennies potential~\cite{TANG1984} of
Ref.~\cite{LI2011} to model the Ba$_2$ $X^1\Sigma^+_g$ potential
as
\begin{equation}
\label{TT}
V(R)=Ae^{-bR}-\sum_{n=3}^{5}\left(1-e^{-bR}\sum_{k=0}^{2n}\frac{(bR)^k}{k!}
\right)\frac{C_{2n}}{R^{2n}},
\end{equation}
where the short-range parameters (in a.u.) $A=105.4$ and
$b=0.9657$ are obtained from rescaling an accurate Sr$_2$
$X^1\Sigma^+_g$ potential~\cite{STEIN2008}, using the measured
fundamental vibration frequency~\cite{LEBEAULT1998}, and taking
the long-range coefficients from Ref.~\cite{PORSEV2002,PORSEV2006}
(see Fig.~\ref{potentials}).

\begin{figure}[t!]
\center
\includegraphics[width = 85 mm, angle = 0]{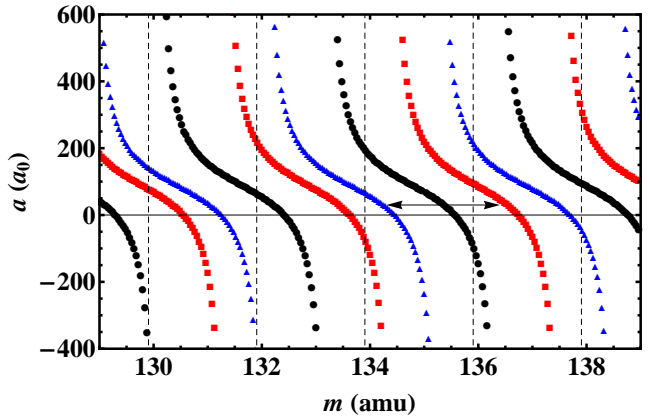}\\
\caption{(Color online) The mass dependence of the scattering
length for Ba, using the Tang-Toennies model to represent the
$X^1\Sigma^+_g$ potential with the parameter values from
Ref.~\cite{LI2011} (black closed circles). The arrow indicates the
effect of the uncertainty in the theoretical $C_6$ coefficient,
showing its lower (blue closed triangles) and upper (red closed
squares) limit. The plotted mass is twice the reduced mass and the
dashed vertical lines indicate the mass of the even isotopes of
Ba.} \label{Bascattering}
\end{figure}

Although we do not expect the potential to be accurate enough to
predict the scattering length, we can use it to investigate its
mass dependence. The result is shown in Fig.~\ref{Bascattering}
(black closed circles). We observe that the spacing between the
resonances is larger than that between the even isotopes, giving
rise to a variety of scattering lengths among the different
isotopes. Therefore it is very probable that at least one of the
isotopes has a favorable scattering length for BEC formation.

In addition, we have investigated the effect of the theoretical
uncertainty in the reported $C_6$ coefficient, $5160\pm74$ a.\ u.\
\cite{PORSEV2002,PORSEV2006}. We have repeated the above
calculation with the upper (red closed squares) and lower (blue
closed triangles) limits~(see Fig.~\ref{Bascattering}). Note that
these adjustments in $C_6$ also require small changes in $A$ and
$b$ \cite{LI2011}. The uncertainty in $C_6$ gives rise to a large
spread in the scattering length, which can also be seen as a large
shift along the mass axis, as indicated by the arrow. This means
that even if an accurate short-range potential would be available,
still no precise prediction of the scattering length is possible.
On the other hand, whereas the scattering length itself is
extremely sensitive to details of the potential, the spacing
between the scattering resonances is not. Therefore, once
scattering length information of one of the isotopes becomes
available, the simple Tang-Toennies potential is sufficient to
predict the scattering lengths of all the other isotopes
\footnote{We have applied the Tang-Toennies potential model for
Ca$_2$ \cite{YANG2009} and Sr$_2$ \cite{LI2011}, from which we
have obtained similar results as in Sec.~\ref{calcium} and
Ref.~\cite{STEIN2010}, respectively, when finetuning the $C_6$
coefficient to reproduce the scattering length of one particular
isotope.}, at least at an accuracy that allows to choose the best
isotope to achieve BEC.

\section{Conclusion and Outlook}

In summary, we have investigated the $s$-wave scattering lengths
of the Ca and Ba even (bosonic) isotopes. An experimentally
accurate ground state potential of Ca$_{2}$ enabled us to
calculate the scattering lengths for all the isotopes, taking
knowledge of $^{40}$Ca scattering length as a reference. We have
observed that accidentally scattering resonances appear near all
even isotopes, leading to large positive and negative scattering
lengths for all isotopes. Only $^{42}$Ca shows a small improvement
in terms of scattering length compared to $^{40}$Ca. Therefore,
alternative methods to that of evaporative cooling are interesting
for Ca, such as direct laser cooling into quantum degeneracy using
a narrow transition~\cite{ADAMS2003}.

In addition, scattering lengths for all the even Ba isotopes have
been investigated using a Tang-Toennies model potential. In
contrast to the Ca case, here the scattering lengths vary strongly
over the isotopes, resulting in a high probability that at least
one isotope has a favorable scattering length to produce a BEC.
More accurate knowledge of the ground state Ba$_2$ potential,
including the long-range coefficients, based on experimental data
that covers the full vibrational spectrum is needed for an
improved determination of the scattering lengths. However, once
experimental knowledge on the scattering length of one isotope
becomes available, the scattering lengths of all the other
isotopes can be predicted with the present model potential.

\begin{acknowledgments}
We thank Erling Riis and Wim Vassen for careful reading of the
manuscript and useful comments. This work has been performed as
part of the research program of the \emph{Stichting voor
Fundamenteel Onderzoek der Materie} (FOM) through programme 114
(TRI$\rm{\mu}$P), which is financially supported by the
\emph{Nederlandse Organisatie voor Wetenschappelijk Onderzoek}
(NWO). S.\ K.\ acknowledges financial support from NWO (VIDI
grant).
\end{acknowledgments}

\end{document}